\documentclass[aps,prb,reprint,superscriptaddress,showpacs]{revtex4-1}
\usepackage{bm}
\usepackage{graphicx}
\begin{document}

\title{Sensitive detection of level anti-crossing spectra
 of nitrogen-vacancy centers in diamond}

\author{S.~V.~Anishchik}
\email[]{svan@kinetics.nsc.ru} \affiliation{Voevodsky Institute of
Chemical Kinetics and Combustion SB RAS, 630090, Novosibirsk,
Russia}

\author{K.~L.~Ivanov}
\email[]{ivanov@tomo.nsc.ru} \affiliation{International Tomography
Center SB RAS, 630090, Novosibirsk, Russia}
\affiliation{Novosibirsk State University, 630090, Novosibirsk,
Russia}

\begin{abstract}
We report a study of the magnetic field dependence of
photoluminescence  of NV$^-$ centers (negatively charged
nitrogen-vacancy centers) in diamond single crystals. In such a
magnetic field dependence characteristic sharp features are
observed, which are coming from Level Anti-Crossings (LACs) in a
coupled electron-nuclear spin system. For sensitive detection of
such LAC-lines we use lock-in detection to measure the
photoluminescence intensity. This experimental technique allows us
to obtain new LAC lines. Additionally, a remarkably strong
dependence of the LAC-lines on the modulation frequency is found.
Specifically, upon decrease of the modulation frequency from 12
kHz to 17 Hz the amplitude of the LAC-lines increases by
approximately two orders of magnitude. To take a quantitative
account for such effects, we present a theoretical model, which
describes the spin dynamics in a coupled electron-nuclear spin
system under the action of an oscillating external magnetic field.
Good agreement between experiments and theory allows us to
conclude that the observed effects are originating from coherent
spin polarization exchange in a coupled spin system comprising the
spin-polarized NV$^-$ center. Our results are of great practical
importance allowing one to optimize the experimental conditions
for probing LAC-derived lines in diamond crystals comprising
NV$^-$ centers and for indirect detection and identification of
other paramagnetic defect centers.
\end{abstract}


\maketitle

\section{Introduction}

The negatively charged nitrogen-vacancy defect center (NV$^-$
center) in diamond  is of great interest due to its unique
properties \cite{Doherty2013}. NV$^-$ centers are promising
systems for numerous applications, in particular, for quantum
information processing
\cite{Gruber1997,Wrachtrup2001,Jelezko2004r,Childress2006,
Wrachtrup2006,Hanson2006b,Gaebel2006,Santori2006o,Waldermann2007,Maurer2012,
vanderSar2012,Dolde2013,Dolde2014,Pfaff2014} and nanoscale
magnetometry
\cite{Taylor2008,Balasubramanian2008,Maze2008,Rittweger2009,
Acosta2009,Fang2013}. It is well-known that upon optical
excitation the triplet ground state of the NV$^-$ center acquires
strong electron spin polarization. Due to magnetic dipole-dipole
interactions between NV$^-$ centers and other paramagnetic defects
in the crystal spin polarization exchange can occur. Such a
polarization transfer is of relevance for many applications
\cite{Maurer2012,Jarmola2015,Mrozek2015,Chen2016}. An informative
method for studying such polarization transfer processes is given
by the Level Anti-Crossing (LAC) spectroscopy. At LACs there is no
energy barrier for polarization transfer; consequently, coupled
spins can efficiently exchange polarization. As usual, by an LAC we mean the
following situation: at a particular field strength a pair of
levels, corresponding to quantum states $|K\rangle$ and
$|L\rangle$, tends to cross but a perturbation $V_{KL}\neq 0$
lifts the degeneracy of the levels so that the crossing is
avoided. It is well-known that at an LAC efficient coherent
exchange of populations of the $|K\rangle$ and $|L\rangle$ states
occurs
\cite{Colegrove1959,Ivanov2014,Pravdivtsev2014,Clevenson2016}.

LACs give rise to sharp lines in the magnetic field dependence  of
the photoluminescence intensity of the NV$^-$ center. The most
pronounced line \cite{Epstein2005} is observed at 1024 G, which
comes from an LAC of the triplet levels in the NV$^-$ center.
Other lines are termed, perhaps, misleadingly, cross-relaxation
lines \cite{VanOort1989}. In reality, all these lines are due to
the coherent spin dynamics caused by spin polarization exchange at
LACs of the entire spin system of interacting defect centers.
Thus, it is reasonable to term the observed magnetic field
dependences ``LAC spectra''.

In this work, we report a study of LAC-lines in diamond single
crystals by using modulation of the external magnetic field.
Generally, LAC-lines are observed by monitoring photoluminescence
as a function of the external magnetic field; a prerequisite for
such experiments
\cite{VanOort1989,Epstein2005,Hanson2006,Rogers2008,Rogers2009,Lai2009,
Armstrong2010,Anishchik2015,Zheng2017} is precise orientation of
the diamond crystal (so that the magnetic field is parallel to
[111] crystal axis with a precision of better than one tenth of a
degree). Typically, the LAC-line at 1024~G is relatively easy to
detect; however, observation of weaker satellite lines coming from
interaction with other paramagnetic centers is technically more
demanding. Generally, the experimental method using low-amplitude
modulation of the external magnetic field and lock-in detection
provides much better sensitivity to weaker sharp lines. In such
experiments the external field strength is modulated at a
frequency $f_m$; the output luminescence signal is multiplied by
the reference signal given by $\cos(2\pi f_mt)$ or $\sin(2\pi
f_mt)$ and integrated over time to provide an increased
sensitivity to weak signals. In experiments using lock-in
detection \cite{Anishchik2015,Anishchik2016b} a new LAC line at
zero magnetic field has been found recently; additionally, groups
of LAC-lines around 5--250~G, 490--540~G, 590~G and 1024~G have
become visible. Some of these lines have been observed
\cite{Anishchik2016b} for the first time; they originate from the
interaction of the NV$^-$ center with other paramagnetic defect
centers in the crystal. Detailed analysis of these lines and
discussion of the defect centers detected by investigating
LAC-lines can be found elsewhere. In this work we focus on the
spin dynamics behind the detection method using field modulation.
It is common that the shape of the lines (``dispersive''
lineshape) obtained with field modulation is different from that
fund without modulation: each line has a positive and a negative
component, at the center of each line the signal intensity is
zero. At first glance, such an appearance of the LAC-lines
(``derivative'' spectrum) is standard for experiments using
lock-in detection. However, here we demonstrate an unexpected
behavior of the LAC-lines, namely, a substantial increase of the
line amplitude upon decrease of the modulation frequency. Such an
increases is crucial for the detection of weak LAC-lines coming
from the interaction of the NV$^-$ center with other defect
centers present only in very small concentration.

\section{Methods}

\subsection{Experimental}

The experimental method is described in detail in a previous
publication \cite{Anishchik2015}.

Experiments were carried out using single crystals of a  synthetic
diamond grown at high temperature and high pressure in a Fe-Ni-C
system. As-grown crystals were irradiated by fast electrons of an
energy of 3~MeV; the irradiation dose was  $10^{18}$~el/cm$^2$.
After that the samples were annealed for two hours in vacuum at a
temperature of 800$\rm ^o$Ñ. The average concentration of NV$^-$
centers was $9.3\times10^{17}$ cm$^{-3}$.

The samples were placed in a magnetic field, which is a
superposition  of the permanent field, $B_0$, and a weak field
modulated at the frequency $f_m$:
\begin{equation}
    B=B_0+B_m \cos(2\pi f_m t), \label{bmod}
\end{equation}
and irradiated by the laser light at a wavelength of 532~nm
(irradiation power was 400 mW). The beam direction was
perpendicular to the magnetic field vector $\bm{B}_0$. The laser
light was linearly polarized and the electric field vector
$\bm{E}$ was perpendicular to $\bm{B}_0$. The luminescence
intensity was measured by a photo-multiplier. The resulting signal
was send to the input of the lock-in detector. The modulation
frequency $f_m$ was varied from 17~Hz to 12.5~kHz.

\subsection{Theory}
Generally, field modulation is a method  providing better
sensitivity to weak and sharp lines; by using modulation one
typically obtains ``derivative'' spectra. Indeed, when the field
dependence of the measured signal is given by a function $S(B)$
for the field given by expression (\ref{bmod}) we obtain the
following signal:
\begin{equation}
    S(B)\approx S(B_0)+B_m \cos(2\pi f_m t)\frac{dS}{dB}. \label{smod}
\end{equation}
This expression is valid for $B_m\ll B_0$. Hence, the
time-dependent contribution to the signal oscillates at the $f_m$
frequency and its amplitude is given by the $dS/dB$. Furthermore,
there is no phase shift between the field modulation and the
signal. However, this simple consideration contradicts to  our
experimental data, necessitating development of a more consistent
approach to the problem under study. Specifically, we need to
treat the spin dynamics induced by the modulated magnetic field.

\begin{figure}
   \includegraphics[width=0.4\textwidth]{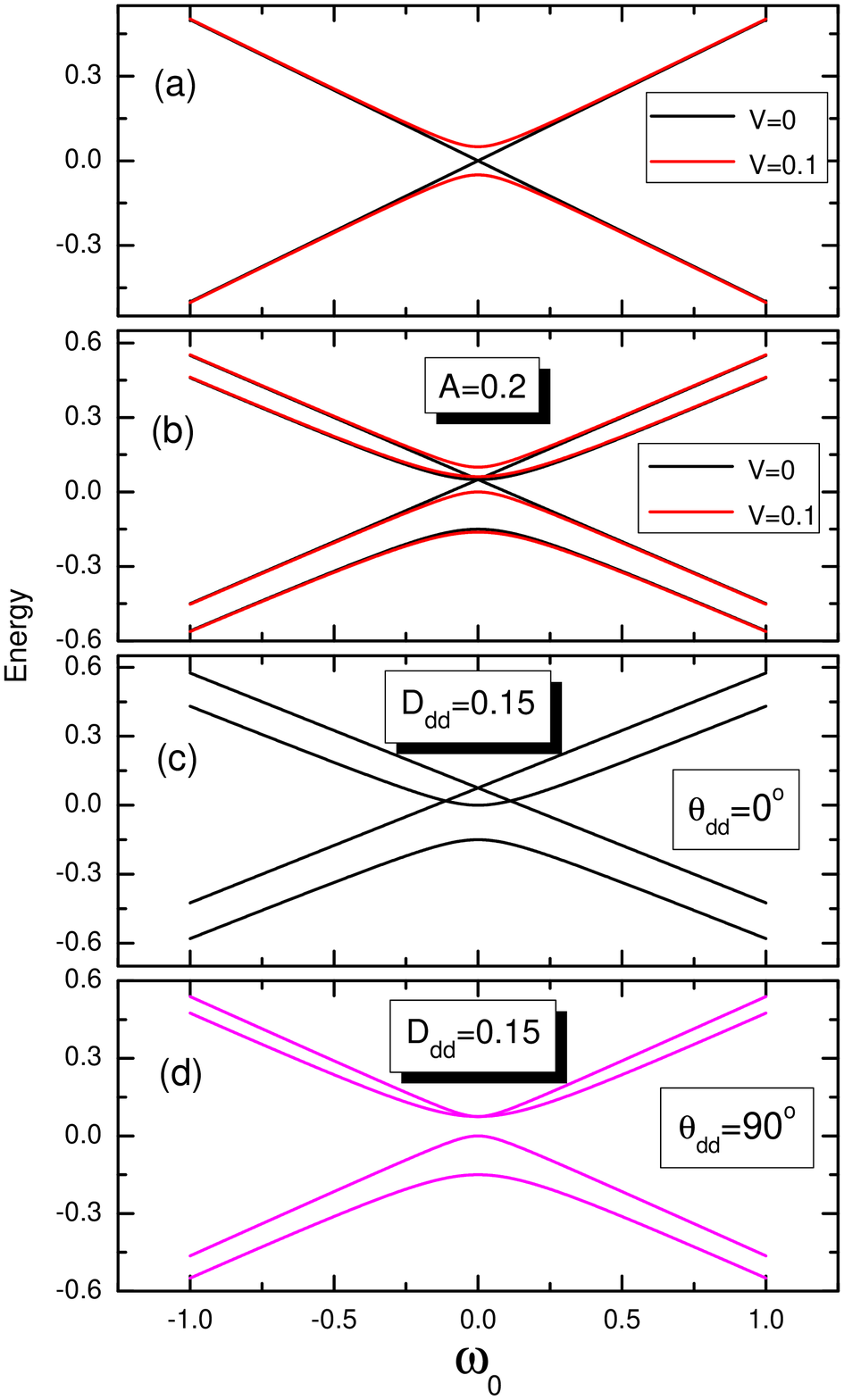}
   \caption{Energy levels of a single-spin (a) and two-spin
   (b--d) model systems. (a) Energy levels of the spin in
    the magnetic field directed along the $z$-axis in the absence
     of an external perturbation ($V=0$)   and in its presence
     ($V=0.1$). (b) Energy levels of an electron-nuclear spin
      system coupled by isotropic HFC with $A=0.2$ assuming $V=0$
       and $V=0.1$.  Energy levels of an electron-nuclear spin
        system coupled by dipolar HFC assuming (c) $\bm{n}~||~z$
 ($\theta_{dd}=0$) and (d) $\bm{n}\perp z$ ($\theta_{dd}=90^o$).
 In subplots (c, d) we have taken $A=V=0$.
   \label{levels}}
\end{figure}

In order to understand the spin dynamics behind our  experiments
we perform numerical simulations. To model polarization transfer
in the electron-nuclear spin system we make the following
simplifications. First, we do not treat the entire three-level
electron spin system but restrict ourselves to only two levels.
Such a simplification is reasonable owing to the sizable
zero-field splitting in the NV$^-$ center. Consequently, only two
triplet sublevels can closely approach each other (at particular
matching conditions) whereas the third level stays far apart from
them. In such a situation the electronic spin subsystem can be
modeled by a fictitious\cite{Feynman1957,Vega1977} spin $S=1/2$.
We also assume that the luminescence intensity is proportional to
the population of the $S_z=1/2$ state, hereafter, the
$\alpha$-state: i.e., the system has a ``bright'' state, which
provides fluorescence, and the ``dark'' $S_z=-1/2$ state,
hereafter, the $\beta$-state. This is a reasonable assumption
because only one of the three triplet states of the NV$^-$ center
gives rise to intense luminescence. Hereafter, we assume that the
$z$-axis is parallel to the external magnetic field. The $S$ spin
interacts with the permanent external $B_0$ field and with the
oscillating $B_m$ field. In this situation, the Hamiltonian of the
spin system is of the form (in $\hbar$ units):
\begin{equation}\label{Ham0}
    \hat{H}(t)=\gamma B_0 \hat{S}_z+V \hat{S}_x+\gamma B_m\cos( 2\pi f_mt)\hat{S}_z,
\end{equation}
where $\hat{\bm{S}}$ is the spin operator of the electron,
$\gamma$ is the  electronic gyromagnetic ratio, $V$ is an external
perturbation (coming, e.g., from a small misalignment of the
crystal). Hereafter we use notations $\gamma B_0=\omega_0$,
$\gamma B_m=\Omega_1$. Considering only the main part of the
Hamiltonian, $H_0=\omega_0 \hat{S}_z$ we obtain that there is a
level crossing at $B_0=0$; however, the perturbation given by $V$
mixes the crossing levels and turn this crossing into an LAC (See
Fig. \ref{levels}a). By turning on the modulation we introduce
repeated passages through the LAC; upon these passages spin
evolution is taking place resulting in redistribution of
polarization.

Here we also extend the treatment to
electron-nuclear spin systems, i.e., we consider interaction of
the electron spin with surrounding nuclear spins by hyperfine
coupling (HFC). For the sake of simplicity, we reduce the nuclear
spin subsystem to only one spin $I=1/2$. Then the Hamiltonian of
the spin system under consideration takes the form (in $\hbar$
units):
\begin{equation}\label{Ham}
    \hat{H}(t)=\omega_0 \hat{S}_z+V \hat{S}_x + A(\hat{\bm{S}}\cdot
    \hat{\bm{I}})+\Omega_1\cos( 2\pi f_mt)\hat{S}_z,
\end{equation}
where $\hat{\bm{I}}$ is the spin operator of the nucleus, $A$ is
the isotropic HFC constant. The energy levels of a two-spin
system are shown in Fig. \ref{levels}b taking account for the
isotropic HFC and the $V$-term. Additionally, we consider a model
where dipolar HFC is used instead of isotropic HFC:
\begin{eqnarray} \label{dip}
 \nonumber \hat{H}(t)= \omega_0 \hat{S}_z+
    D_{dd}[3(\hat{\bm{S}},\bm{n})(\hat{\bm{I}},\bm{n})-
    (\hat{\bm{S}},\hat{\bm{I}})]+ \\
  +\Omega_1\cos( 2\pi f_mt)\hat{S}_z,
\end{eqnarray}
where $D_{dd}$ is the dipolar interaction strength  depending on
the distance between the spins and $\bm{n}$ is the vector pointing
from one spin to the other with unity length, $|\bm{n}|=1$. In
Fig. \ref{levels}c,d we show the energy levels of such a system
for different directions of $\bm{n}$. Here $\theta_{dd}$ is the
angle between $\bm{n}$ and the $z$-axis, which is parallel to the
external magnetic field. Hereafter, for the sake of simplicity,
all parameters of the spin Hamiltonian as well as spin relaxation
parameters are given in dimensionless units.

In our model, the observable signal is given by the population  of
one of the states of the $S$-spin, for clarity, the bright state
is the $\alpha$-state. To make comparison with the experiments we
multiply the  population of the $\alpha$-state,
$\rho_{\alpha\alpha}$, by the $\cos(2\pi f_mt)$ function and
integrate it over the modulation period, see below.

\begin{figure}
   \includegraphics[width=0.4\textwidth]{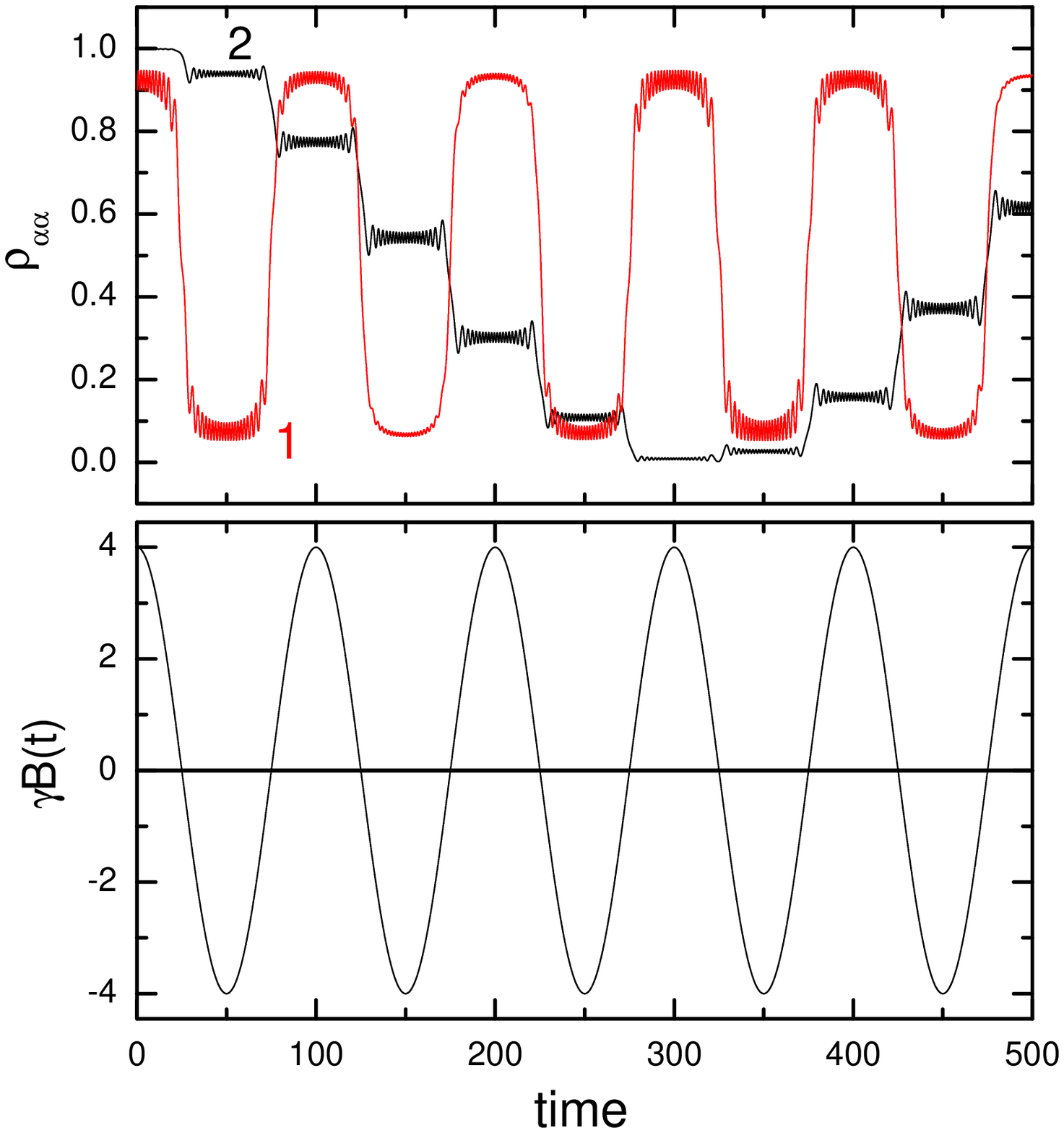}
   \caption{Spin dynamics resulting from passage through the
   LAC in a two-level system described by the Hamiltonian $\hat{H}(t)$
   given by eq. (\ref{Ham0}). Top: population of the $\alpha$-state
   for adiabatic (curve 1) and non-adiabatic (curve 2) variation of the
   Hamiltonian. Bottom: magnetic field $\gamma B(t)=\Omega_1 \cos(2\pi f_mt)$.
   Calculation parameters: $\omega_0=0$, $\Omega_1=4$, $f_m=0.01$,
   $V=1$ (curve 1) and 0.1 (curve 2); at $t=0$ the system is in the
    ``bright'' state; relaxation
    effects are neglected.\label{spindyn}}
\end{figure}

Qualitatively, we expect different regimes for spin dynamics at
$f_m\ll V$ and $f_m\gg V$ as demonstrated in Fig.~\ref{spindyn} for a
two-level system. At $f_m\ll V$ each passage through the LAC
results in adiabatic inversion of populations of the $S$-spin
states. Consequently, the luminescence signal is expected to be
modulated at the $f_m$ frequency having the maximal possible
amplitude and the same phase as the modulated external field.
During a fast passage through the LAC, i.e., at $f_m\gg V$, the
populations are mixed only slightly in each passage and the
amplitude and frequency of modulation of the luminescence signal is expected to
drop down. In addition, modulation of the signal is no longer
in-phase with the reference signal of the lock-in amplifier,
resulting in both considerable phase shifts and reduction of the signal. As
we show below, the calculation results are in good agreement with
this simple consideration.

For systematic analysis we also take spin relaxation into
account. To do so, we treat the spin evolution as described by the
Liouville-von Neumann equation:
\begin{equation}\label{Liu1}
    \frac{d\rho_{ij}}{dt}=L_{ij;kl}(t)\rho_{kl},
\end{equation}
where $\rho$ is the density matrix of the two-spin
electron-nuclear   system in the Liouville representation (column
vector with 16 elements), while the elements of the
$\hat{\hat{L}}$ super-operator are as follows:
\begin{equation}\label{Liu2}
    L_{ij;kl}=i(\delta_{ik}H_{lj}-\delta_{jl}H_{ik})+R_{ij;kl},
\end{equation}
where $R_{ij;kl}$ is the relaxation matrix. To specify the $R$
super-operator  we make the following simplifying assumptions. We
treat two contributions to the electron spin relaxation, the
longitudinal relaxation  (relaxation of  populations having the
rate $R_1$) and transverse relaxation (relaxation of
coherences having the rate $R_2$). We consider two different
cases: $R_1=R_2$ and $R_1\ll R_2$, see below. In addition, we take
into account photo-excitation of the NV$^-$ center, which produces
the electron spin polarization, i.e., the population difference
for the states of the $S$-spin. This process is considered in a
simplified manner as a transition from the $\beta$-state to the
$\alpha$-state at a rate $J$ (pumping rate for the electron spin
polarization). Thus, for the sake of simplicity, we do not
consider the complete excitation cycle in the NV$^-$ center, which
gives rise to the electron spin polarization. Relaxation of the
nuclear spin is completely neglected because it is usually much
slower than that for the electron spin.

\begin{figure}
   \includegraphics[width=0.4\textwidth]{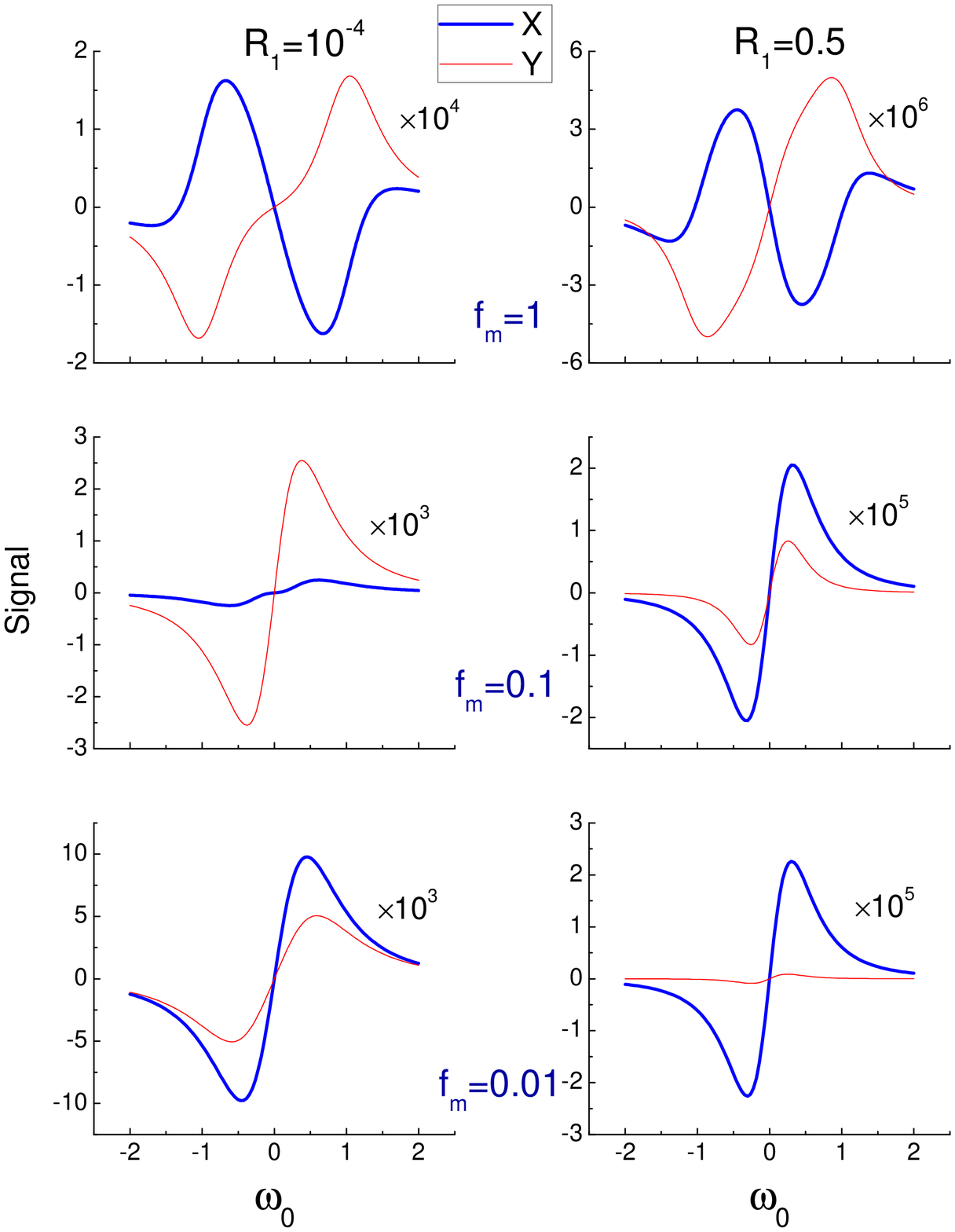}
   \caption{Theoretical LAC-spectra of a single-spin system for
   three different $f_m$ frequencies
    equal to 1 (top), 0.1 (middle) and 0.01 (bottom) and
   longitudinal relaxation rate $R_1$ equal to $10^{-4}$
   (left) and $0.5$ (right). Here both components of the signal,
   $X$ (thick lines) and $Y$ (thin lines), are shown, calculated
   according to eqs. (\ref{XY1}) and (\ref{XY2}), respectively.
   Other calculation parameters are: $R_2 = 0.5$,  $J= 0.01$, $V = 0.1$,
$\Omega_1=0.1$.\label{spectra}}
\end{figure}

To perform numerical calculations we split the modulation period
$T=1/f_m$  into $N$ equal intervals  of a duration $\Delta t=T/N$.
In each step, the density matrix $\rho$ was propagated by using a
matrix exponent:
\begin{equation}\label{Liu3}    \rho(t+\Delta
t)=\exp[\hat{\hat{L}}(t)\Delta t]\rho(t).
\end{equation}
Generally, the solution depends on the initial conditions.
However, in  the present case we are interested in the
``steady-state'' solution, which is reached after many modulation
periods. Indeed, in experiments transient effects are not
important because signal averaging is performed over many $T$
periods (only the steady-state of the system is probed). To obtain
such a solution of eq. (\ref{Liu1}) we assume that the density
matrix before a period of modulation, $\rho(0)$, is the same as
that after the period:
\begin{eqnarray}
\nonumber  \rho(0)=\rho(T) &=& \exp[\hat{\hat{L}}(t=T-\Delta t)\Delta t]\times\ldots \\
  ~ &~& \times\exp[\hat{\hat{L}}(t=0)\Delta t]\rho(0)=\hat{\hat{U}}\rho(0).
  \end{eqnarray}
Here $\hat{\hat{U}}$ is the super-matrix, which describes  the
evolution over a single modulation period. This equation is a
linear equation for the $\rho(0)$ vector. To find a non-trivial
solution of such a matrix equation, we need to exclude one
equation from the system (the one, which linearly depends on other
equations) and to replace it by the expression
$\sum_i\rho_{ii}(0)=1$, which describes nothing else but
conservation of the trace of the density matrix. This new system
can be solved by using linear algebra methods. The $\hat{\hat{U}}$
matrix is computed numerically; to do so we set the value of $N$
such that further increase of $N$ changed the final result by less
than 1\%. Of course, it is necessary to increase $N$ substantially
at small $f_m$. At the lowest modulation frequency we typically use
$N=2\times 10^6$.

To compare theoretical results to the experimental data we
numerically compute the sine and cosine Fourier components of the
element of interest of the density matrix, namely, the population
of the $\alpha$-state, $\rho_{\alpha\alpha}$. This element can be
computed when $\rho(t)$ is known:
\begin{equation}\label{XY1}
    X=\frac{1}{T}\int_0^{T}\rho_{\alpha\alpha}\cos(2\pi f_mt)dt,
\end{equation}
\begin{equation}\label{XY2}
    Y=\frac{1}{T}\int_0^{T}\rho_{\alpha\alpha}\sin(2\pi f_mt)dt.
\end{equation}
Knowing $X$ and $Y$ we can completely characterize the signal. An
analogue of the lock-in detector phase variation by an angle
$\phi$ is the rotation of axes in the functional space:
    \begin{equation}\label{newx}
    X'=X\cos\phi + Y\sin\phi.
    \end{equation}

Typical calculated LAC-spectra are presented in Fig. \ref{spectra}
for three different $f_m$ frequencies; both components of the
signal, $X$ and $Y$, are shown. Calculation of the two signal
components is performed using eqs.  (\ref{XY1}) and (\ref{XY2}).
When the modulation frequency is low and relaxation of populations
is relatively fast the $X$-component of the signal is strong
whereas the $Y$-component is negligible, i.e., the signal is
cosine-modulated and there is virtually no phase shift with
respect to the modulated input signal $\Omega_1\cos(2\pi f_mt)$.
At higher modulation frequency both components of the signal are
significant, i.e., there is a strong phase shift with respect to
the input signal. The appearance of the spectra changes upon
variation of the $R_1$ rate. Upon decrease of the $f_m$ frequency
not only the $Y$-component is reduced but also the signal
intensity grows. As we demonstrate below, such a behavior of the
LAC-lines is consistent with experimental findings.

\section{Results and Discussion}

\begin{figure}
   \includegraphics[width=0.47\textwidth]{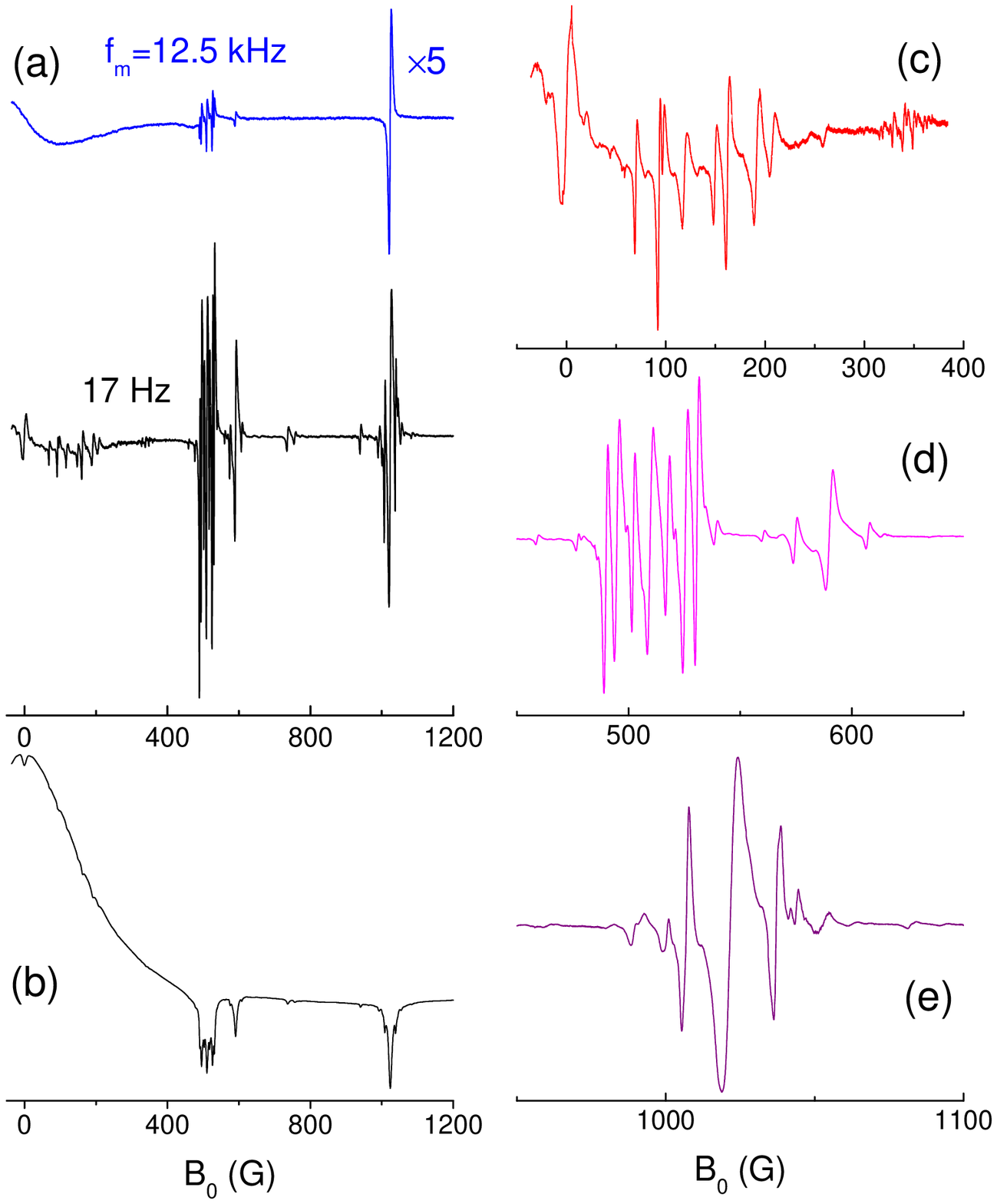}
   \caption{Experimental LAC spectra of NV$^-$ centers in a
    diamond single crystal. (a) LAC spectra in the range
    $-50$-$1200$~G obtained using the modulation frequency $f_m$
     of 12.5 kHz and 17 Hz; the
      upper trace is multiplied by 5. (b) Integrated LAC spectrum
      obtained with $f_m=17$ Hz.
      Subplots (c), (d) and  (e) show enlarged different regions of the
       LAC spectrum. In all cases the modulation amplitude was
       $\Omega_1=0.5$ G. Except for the upper trace in (a) the
        modulation frequency was 17 Hz. The lock-in detector phase
         was chosen such that the LAC-line at 1024 G has maximal
          amplitude. \label{fig1}}
    \end{figure}

\subsection{Experimental LAC-spectra}

In Fig. \ref{fig1} we show the transformation of  the LAC-spectra
upon variation of the modulation frequency. In subplot (a) we show
the results for two different frequencies, 12.5~kHz and 17~Hz. One
can readily see that the amplitude of all lines is much higher at
the low $f_m$ value. For better presentation we compare the
original spectra with integrated LAC spectra: in the integrated
spectra all LAC-lines show up as dips in the $B_0$ dependence. One
can readily see that LAC-lines are much better visible in the
spectra obtained with field modulation. LAC-lines  that show up
only at low modulation frequency are analyzed in our previous work
\cite{Anishchik2016b} and originate from polarization transfer
between paramagnetic defect centers. Here we only briefly mention
the main peculiarities of the detected LAC-lines, see Fig.
\ref{fig1}c,d,e. Polarization transfer between defect centers
occurs when the level splittings in the two centers become equal
to each other (causing a level crossing): under such conditions electronic
dipole-dipole interaction turns a level crossing into an LAC and
enables coherent polarization exchange. Such a polarization
transfer is usually termed (perhaps, erroneously because
polarization transfer is due to a coherent mechanism)
cross-relaxation \cite{VanOort1989}.

The line at zero field\cite{Anishchik2015} comes from polarization
transfer  between two NV$^-$ centers; for symmetry reasons at zero
field energy matching for two NV$^-$ centers always occurs, consequently,
the zero-field line has been found in all samples we studied so
far. Other lines emerging in the field range 50--400 G, which are
visible only at low modulation frequencies, are due to interaction
of the NV$^-$ center with other paramagnetic defect centers in the
crystal; quantitative analysis of the positions and amplitudes of
these lines thus provides valuable information about the Electron Paramagnetic Resonance (EPR)
parameters (spin value, zero-field splitting, hyperfine
couplings) of these centers. The LAC-lines found around 500~G and
at 590~G become considerably stronger at $f_m=17$~Hz; the former
are coming from interaction of the NV$^-$ center with the
P1-center (neutral nitrogen atom, replacing carbon in the diamond
lattice) while the latter comes from interaction of two NV$^-$
centers having different orientations with respect to the external
magnetic field. Additionally, the line at 1024~G corresponds to
the LAC in the NV$^-$ in its ground state. Finally, the satellite
lines at 1007~G and 1037~G are seen in the LAC-spectrum. These
lines are originating from polarization exchange between the
spin-polarized NV$^-$ center and the P1-center\cite{Armstrong2010}.
The low amplitude of the satellite lines is due to the weak
interaction between different defect centers.

Hence, one can see that the amplitude of all LAC-lines is
very sensitive to the $f_m$ value, so that some lines can be found
only using very low modulation frequency. For this reason, we find
it important to analyze the $f_m$ dependence of LAC-lines and
elucidate the parameters that determine this dependence.

\begin{figure}
   \includegraphics[width=0.4\textwidth]{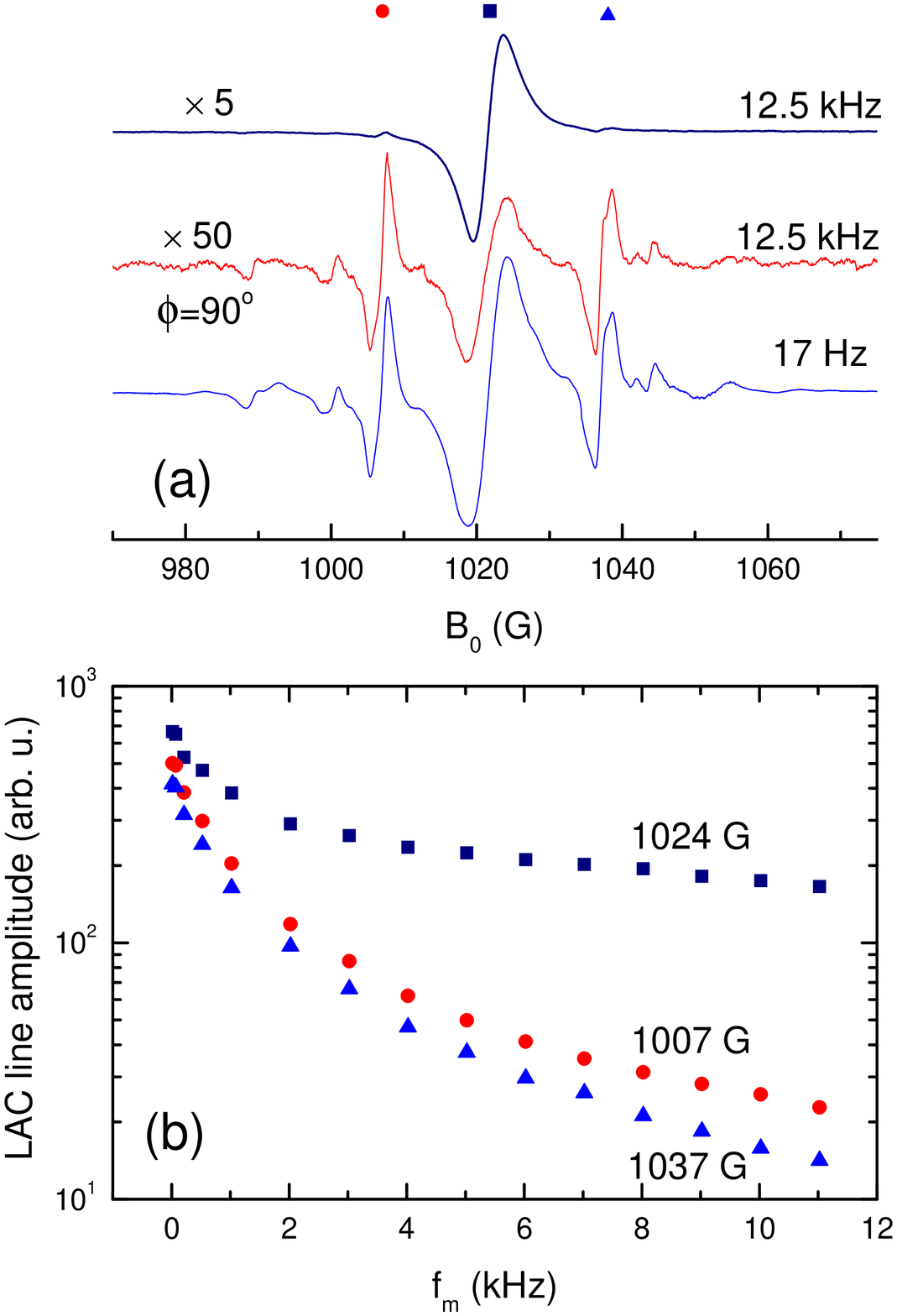}
   \caption{(a) Experimental LAC spectra of NV$^-$ centers in a
    diamond single crystal in the magnetic field range 970-1075 G.
     For each curve the $f_m$ value used in experiments is specified.
      For the upper curve the phase of the lock-in detector is chosen
       such that the signal for the central LAC-line is maximal.
        For the middle trace the phase is shifted by $90^\circ$
         with respect to that for the upper curve. The amplitude
          of the upper curve is increased by a factor of 5, for
           the middle curve -- by a factor of 50. The LAC-lines
            are indicated by circle, square and triangle.
            (b) $f_m$ dependence of the amplitude of the three
             LAC-lines (symbols correspond to the LAC-lines in
              subplot a). For each curve the magnetic field
               strength $B_0$ corresponding to the center of
                the corresponding line is specified. For each
                 experimental point the lock-in detector phase
                  is set such that the amplitude of the
                  corresponding line was maximal. In all
                   cases the modulation amplitude was $B_m=0.5$
                    G. \label{fig2}}
\end{figure}

Fig. \ref{fig2}a shows the LAC-spectra of the NV$^-$ center in the
field range 970-1075 G. The spectra shown for two different
modulation frequencies, 12.5~kHz and 17~Hz, are remarkably
different. As it is seen from the Figure, when $f_m=12.5$ kHz and
the lock-in detector phase is set such that the central line at
1024~G has the maximal amplitude, the satellite lines at 1007~G
and 1037~G are barely visible. When the modulation frequency is
reduced to 17~Hz the amplitude of the central line increases by a
factor of 7, whereas the satellite lines become 50 times stronger. Additionally, at low frequency the phase shift for all lines is negligible in contrast to that at the high modulation frequency. As it is seen from the LAC-spectrum there are no new lines
appearing in the spectrum in this field range but the
signal-to-noise ratio is substantially increased.

\begin{figure}
   \includegraphics[width=0.45\textwidth]{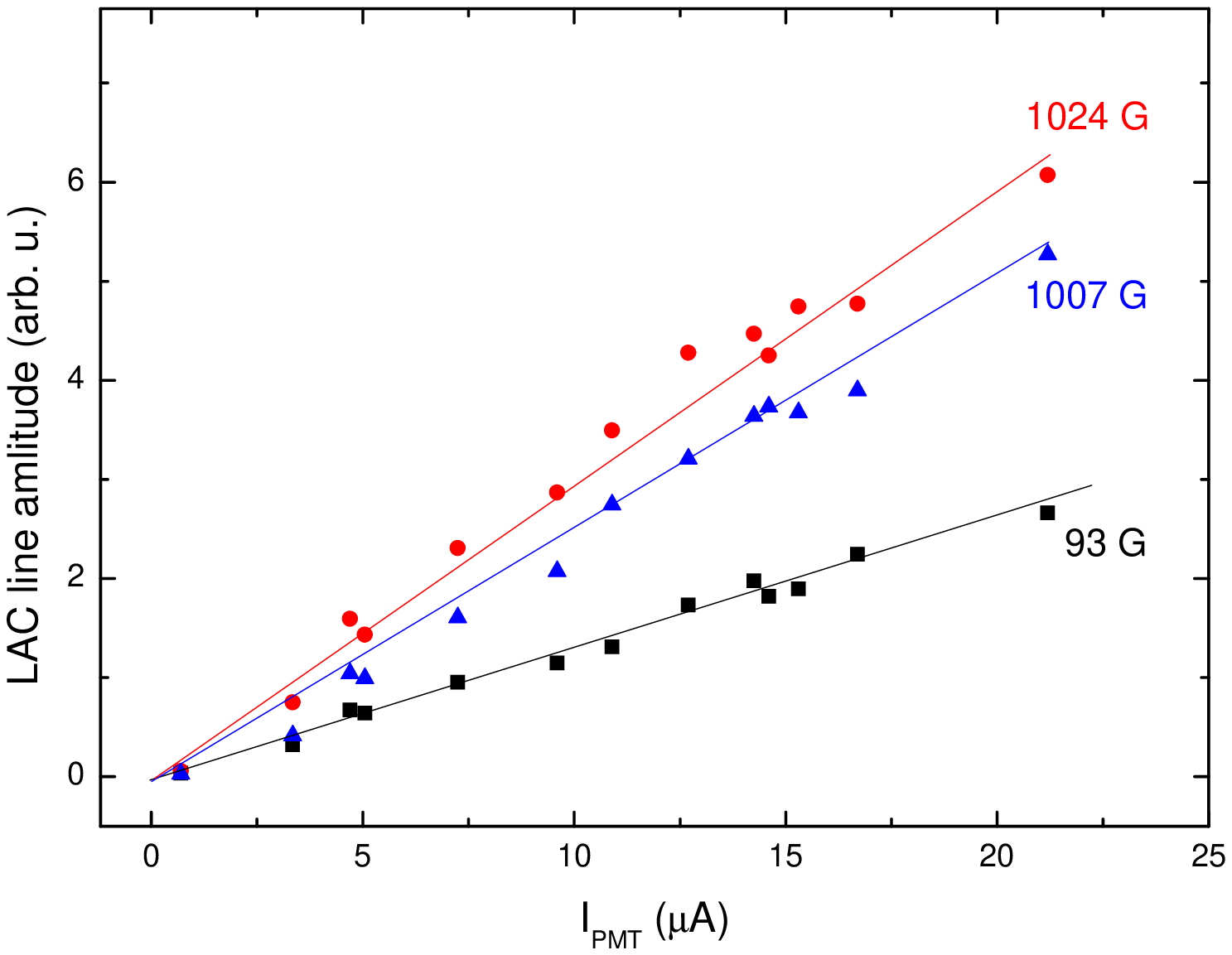}
   \caption{Dependence of the LAC-line at 95~G, 1007~G and 1024~G amplitudes on the luminescence
    intensity of the sample, which is proportional to the current, $I_{PMT}$, in the photo-multiplier. The modulation frequency
   $f_m$ is 17~Hz; the modulation amplitude is 0.5~G.\label{fig5}}
    \end{figure}

In Fig. \ref{fig2}b we present the experimental dependence  of
the line amplitudes, as determined for the three different
LAC-lines, on the modulation frequency. Here the total
peak-to-peak amplitude is presented; the lock-in detector phase
is set such that for each experiment the amplitude of the
corresponding line is maximal. It is clearly seen that by varying
the modulation frequency we obtain a strong variation of the
LAC-line amplitudes, by roughly two orders of magnitude. In the
frequency range under study the dependence is concave, i.e., the slope of
the curve increases at lower modulation frequencies. Hence, by
using modulation we not only obtain the ``derivative'' spectrum:
modulation strongly affects the line amplitudes and shapes. We
attribute this dependence to the spin dynamics caused by
modulation. The most unexpected effect is that the increase of the
line amplitude is occurring at modulation frequencies, which are
much smaller than the electron spin phase relaxation rates of the
NV$^-$ center (when measured in the same units). In samples like
the one we use the phase relaxation times are typically about
several microseconds \cite{Kennedy2003,Rondin2014}. Our estimates
for the phase relaxation times (as determined from the widths of
the EPR and optically detected magnetic resonance spectra) in our samples agree with these values.
Such relaxation times correspond to the frequencies of the order
of several 100 kHz to several MHz. For this reason the growth of
the LAC-line amplitude upon the decrease of $f_m$ from some 10 kHz
to several Hz (in particular, the sharp increase of the line
amplitude at very low frequencies) is perplexing.

In Fig.~\ref{fig5} we show the experimental dependence of the
LAC-line amplitudes on the luminescence intensity of the studied
sample (the current of the photo-multiplier is proportional to
this intensity). Since the luminescence intensity is directly
proportional to the intensity of the excitation light these data
present the dependence of the LAC-line amplitudes on the intensity
of incident light. One can clearly see that the dependence is
almost perfectly linear (in contrast to the quadratic dependence
reported for the zero-field LAC-line \cite{Anishchik2015}). Hence,
the increase of the LAC-line amplitude at low $f_m$ frequencies
cannot be attributed to any two-photon processes.

\subsection{Theoretical calculations}

To rationalize the behavior of the LAC-lines upon variation of
$f_m$  we perform simulations of the spin dynamics of
single-spin and two-spin systems at LACs. Simulations for the
single $S$-spin model properly reproduce the $f_m$-dependence at
high frequencies: the signal decays roughly as $1/f_m^2$. Assuming
$R_1=R_2$ we are not able to reproduce the experimental
dependence: at low $f_m$ the theoretical curve flattens and the
line amplitude does not increase further.

\begin{figure}
   \includegraphics[width=0.4\textwidth]{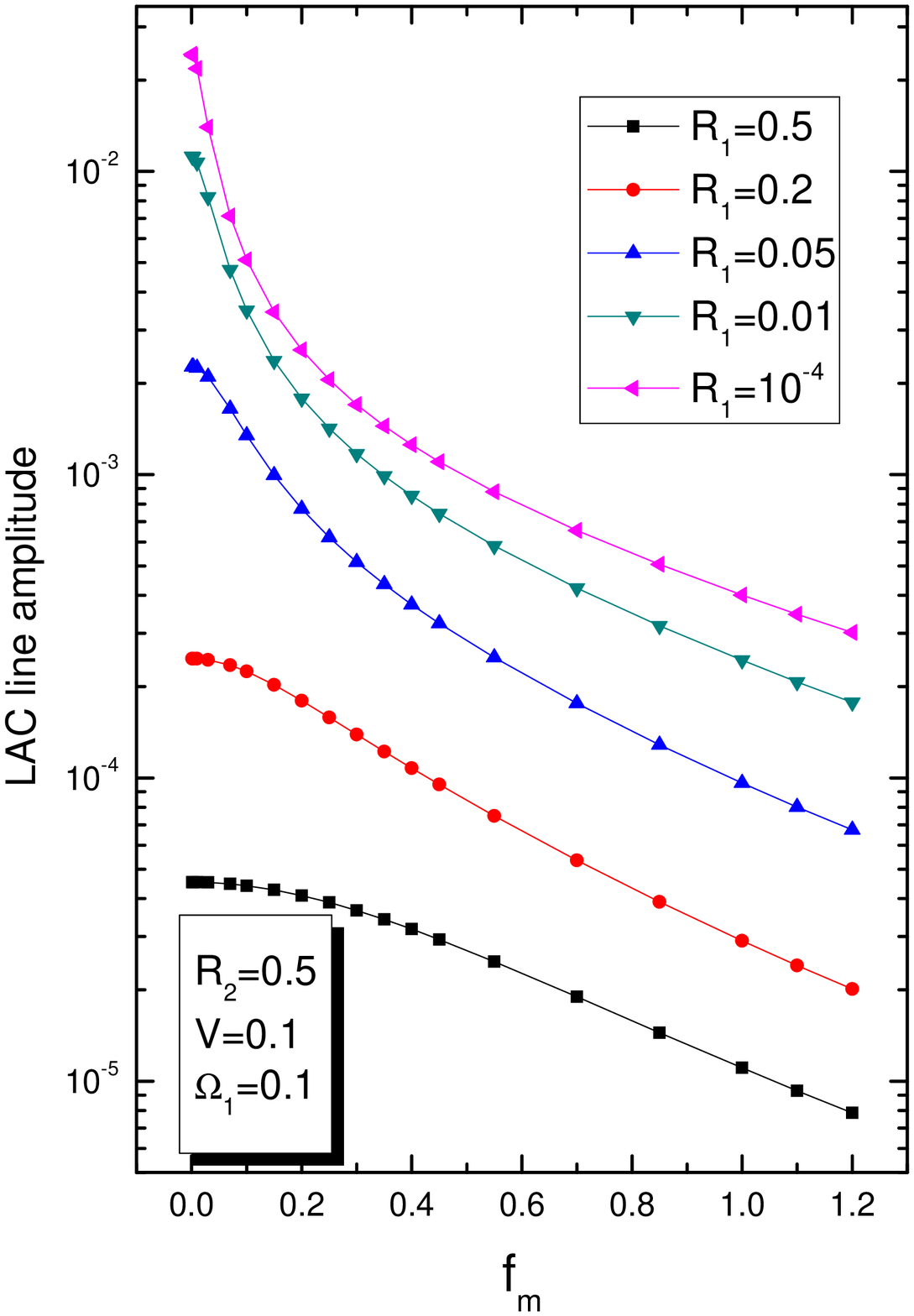}
   \caption{Calculated amplitude of the LAC-line (peak-to-peak)
    for the single $S$-spin model as a function of $f_m$. The
     calculation parameters are given in the graph. \label{fig3}}
\end{figure}

The experimentally observed growth at low frequency is clearly  an
indication of a slow dynamic process in the system. We attribute
this process to $T_1$-relaxation (longitudinal relaxation): in
solids $T_1$-relaxation is usually much slower than
$T_2$-relaxation, i.e., $R_1\ll R_2$. This is correct for the
NV$^-$ centers in diamond as well\cite{Jarmola2012,Jarmola2015}.
Under the assumption $R_1\ll R_2$ we obtain an $f_m$-dependence,
which is much closer to the experimental one: while the behavior
at high $f_m$ remains the same, the LAC-line amplitude continues
to grow at low frequency. It is worth noting that in the
logarithmic-linear coordinates at small $R_1$ the $f_m$ dependence
has the same shape (concave shape, i.e., the second derivative of
the curve is positive; there is an inflexion point at low $f_m$)
as the experimentally observed dependence. However, such a model
predicts a smaller effect of $f_m$ than that found experimentally.
Specifically, at small $f_m$ the calculated curve levels off (the
slope of the curve tends to zero at $f_m\to 0$).

\begin{figure}
   \includegraphics[width=0.4\textwidth]{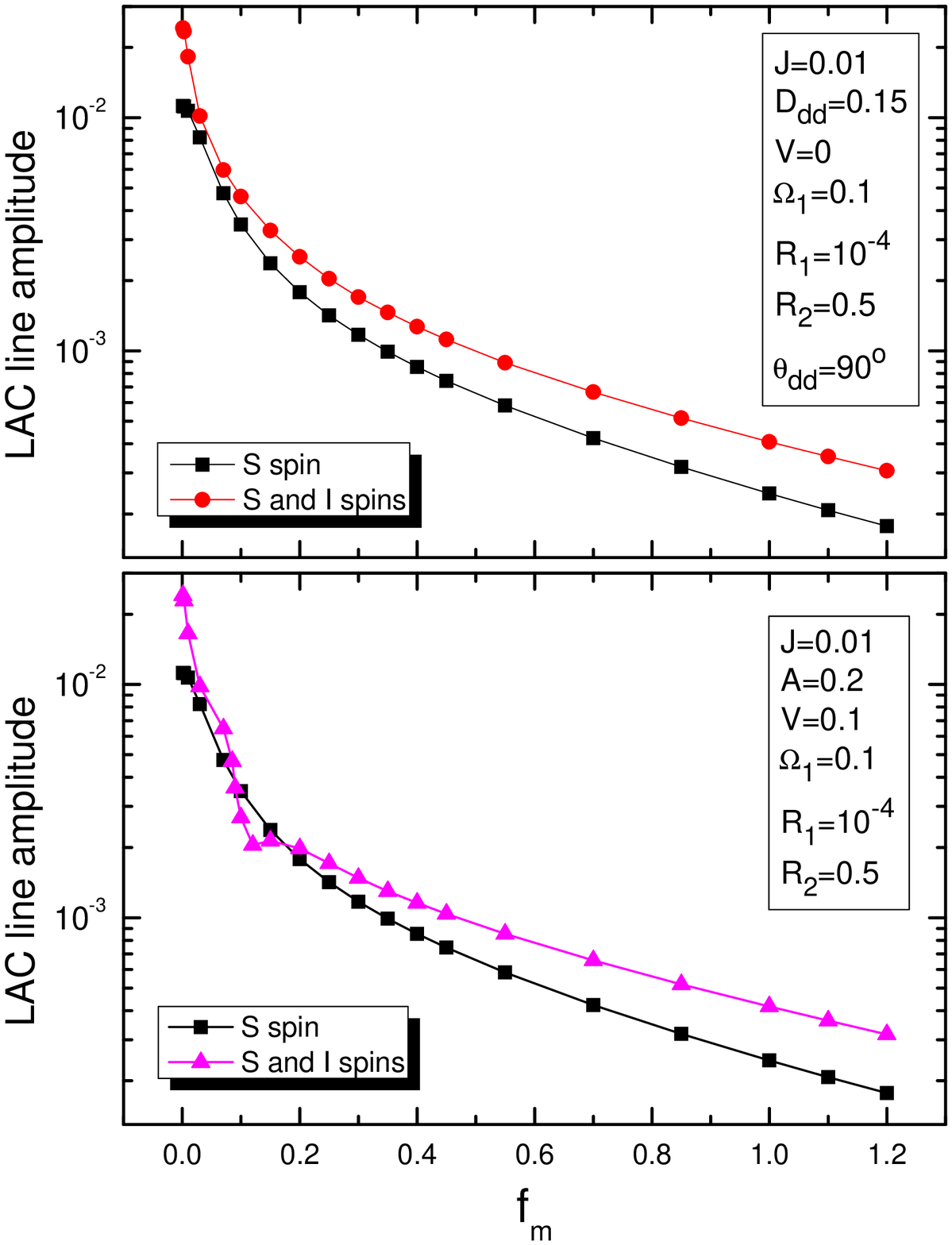}
   \caption{Calculated amplitude of the LAC-line (peak-to-peak)
    for the electron-nuclear two-spin system as a function of $f_m$.
    In the upper graph we present the calculation result for dipolar
    HFC and in the lower graph - for isotropic HFC.
     The calculation parameters are given in the graph.\label{fig4}}
     \end{figure}

The observed strong $f_m$ dependence can be explained by polarization
transfer from the electronic spin system to nuclear spins having
even longer relaxation times than $T_1$. The calculation performed
for the electron-nuclear spin system confirms this expectation:
while at high $f_m$ frequencies the calculation result is almost
the same for the single-spin system and two-spin system, at low
$f_m$ values a sharper increase of the LAC-line amplitude is found
for the electron-nuclear spin system. We attribute this sharper
increase to the effect of the long nuclear $T_1$-relaxation time.

In Fig.~\ref{fig4} we present the calculated $f_m$ dependence of the
LAC-line amplitude for the electron-nuclear spin system assuming
dipolar (upper graph) and isotropic (upper graph) HFC and compare
it to the same dependence for the two-level system. One can see
that in the two-spin system the LAC-line continues to grow even at
low $f_m$; this is due to polarization transfer between the electron
and nucleus. The only difference between the two cases, dipolar
vs. isotropic HFC, is that the curve is monotonous for the dipolar
HFC (similar to the experimental observation), whereas for the
isotropic HFC there is a feature seen at $f_m$ close to the $A$
value. It is also worth noting that assuming isotropic HFC we can
obtain the LAC-line only when $V\neq0$ while for the dipolar HFC
the LAC-line amplitude is non-zero even when $V=0$ (except for the
case $\bm{n}~||~z$).

\section{Conclusions}
We report a study of LAC-lines in the NV$^-$ defect centers in diamond
crystals by using lock-in detection of the signal. Such a method
allows one to obtain sharp LAC-lines with excellent
signal-to-noise ratio. A strong and unexpected effect of the
modulation frequency on the LAC-line amplitude is demonstrated.
Importantly, the LAC-lines are the strongest at low modulation
frequencies. Thus, measurements at low $f_m$ are advantageous,
even despite the technical issues concerning experiments at low
frequencies (namely, the instrumental noise). Moreover,
LAC-spectra obtained at low modulation frequencies are free from
distortions and phase shifts of the signal with respect to the
reference signal of the lock-in amplifier.

To rationalize the observed effect of the modulation  frequency
we performed a theoretical study and computed numerically the
evolution of the spin system under the action of the modulation
field. In the theoretical model, we introduced a single electron
spin 1/2 (modeling the electron spin degrees of freedom of the
NV$^-$ center) coupled to a nuclear spin 1/2. Such a model can
reproduce the main features found in experiments suggesting that
field modulation strongly modifies the dynamics of the spin system
at LACs. Specifically, at low modulation frequency we obtained
adiabatic exchange of populations of the states having an LAC,
whereas non-adiabatic population exchange at high $f_m$ values
leads to decrease of the LAC-line amplitude accompanied by the
phase shift. In order to reproduce the further increase of the
LAC-line amplitude at low $f_m$ frequencies we considered slow
dynamic processes, such as electronic $T_1$-relaxation (which is,
in solids, commonly much slower than $T_2$-relaxation) and nuclear
spin relaxation. Account of these relaxation processes
allowed us to reproduce the experimentally observed $f_m$
dependences.

Our work provides useful practical recommendations on how to
conduct experimental studies of LAC-lines. As we show in a
subsequent publication, the experimental method used here indeed
enables sensitive detection of LAC-lines. Furthermore, for the
first time we demonstrate that modulation (used in lock-in
detection) is not only a prerequisite for sensitive detection of
weak signals but also a method to affect spin dynamics of the
NV$^-$ centers in diamonds. Last but not least, our experimental
method allows one to detect new LAC-lines. Such LAC-lines can be
used for indirect detection of otherwise ``invisible''
paramagnetic defect centers in diamond crystals; their analysis
has been in part performed in Ref. \cite{Anishchik2016b}. A more
detailed analysis will be elsewhere.

\begin{acknowledgments}

Experimental work was supported by the Russian Foundation for
Basic Research (Grant No. 16-03-00672); theoretical work was
supported by the Russian Science Foundation (grant No.
15-13-20035).

\end{acknowledgments}

\bibliography{Anishchik}

\end{document}